\documentclass[aps,twocolumn,superscriptaddress,prl]{revtex4}
\usepackage{amssymb}
\usepackage{amsmath}
\usepackage{graphicx}

\def\gq{e^2/h}

\def\nuone{\nu_1}
\def\nutwo{\nu_2}
\def\alox{\mathrm{Al}_2 \mathrm{O}_3}

\def\VTG{V_{TG}}
\def\VBG{V_{BG}}
\def\He3{^3\mathrm{He}}

\begin{document}
\title{Quantum Hall Effect in a Gate-Controlled p-n Junction in Graphene}
\author{J.\ R.\ Williams}
\affiliation{School of Engineering and Applied Science, Harvard University, Cambridge, MA 02138, USA}
\author{L.\ DiCarlo}
\affiliation{Department of Physics, Harvard University, Cambridge, MA 02138, USA}
\author{C.\ M.\ Marcus}
\affiliation{Department of Physics, Harvard University, Cambridge, MA 02138, USA}

\date{\today}

\begin{abstract}
The unique band structure of graphene allows reconfigurable
electric-field control of carrier type and density, making graphene
an ideal candidate for bipolar nanoelectronics. We report on the
realization of a single-layer graphene \emph{p-n} junction in which
carrier type and density in two adjacent regions are locally
controlled by electrostatic gating. Transport measurements in the
quantum Hall regime reveal new plateaus of two-terminal conductance
across the junction at $1$ and $3/2$ times the quantum of
conductance, $e^2/h$, consistent with recent theory. Beyond enabling
investigations in condensed matter physics, the local-gating
technique demonstrated here sets the foundation for a future
graphene-based bipolar technology.
\end{abstract}

\maketitle
Graphene, a single-layer hexagonal lattice of carbon
atoms, has recently emerged as a fascinating system for fundamental
studies in condensed matter physics~\cite{GeimReview}, as well as a
candidate for novel sensors~\cite{Schedin06,Hwang06} and
post-silicon
electronics~\cite{Berger06,Chen07,Han07,Rycerz07,Lemme07,Huard07,Novoselov04}.
The unusual band structure of single-layer graphene makes it a
zero-gap semiconductor with a linear (photon-like) energy-momentum
relation near the points where valence and conduction bands meet.
Carrier type---electron-like or hole-like---and density can be
controlled using the electric-field effect\cite{Novoselov04},
obviating conventional semiconductor doping, for instance via ion
implantation. This feature, doping via local gates, would allow
graphene-based bipolar technology---devices comprising junctions
between hole-like and electron-like regions, or $p$-$n$
junctions---to be reconfigurable, using only gate voltages to
distinguish $p$ (hole-like) and $n$ (electron-like) regions within a
single sheet. While global control of carrier type and density in
graphene using a single back gate has been investigated by several
groups~\cite{Novoselov05,Zhang05,Heersche07}, local
control~\cite{Lemme07,Huard07} of single-layer graphene has remained
an important technological milestone. In addition, $p$-$n$ junctions
are of great interest for low-dimensional condensed matter physics.
For instance, recent theory predicts that a local step in potential
would allow solid-state realizations of relativistic (``Klein'')
tunneling~\cite{Katsnelson06,Cheianov06}, and a surprising
scattering effect known as Veselago lensing~\cite{Cheianov07},
comparable to scattering of electromagnetic waves in negative-index
materials~\cite{Smith04}.

We report on the realization of local top gating in a single-layer
graphene device which, combined with global back gating, allows
individual control of carrier type and density in adjacent regions
of a single atomic layer. Transport measurements at zero
perpendicular magnetic field $B$ and in the quantum Hall (QH) regime
demonstrate that the functionalized aluminum oxide ($\alox$)
separating the graphene from the top gate does not significantly
dope the layer nor affect its low-frequency transport properties. We
study the QH signature of the graphene $p$-$n$ junction, finding new
conductance plateaus at 1 and $3/2~\gq$, consistent with recent
theory addressing equilibration of edge states at the $p$-$n$
interface~\cite{Abanin07}.

\begin{figure}[b]
\label{fig1.pdf}
\includegraphics[width=3in]{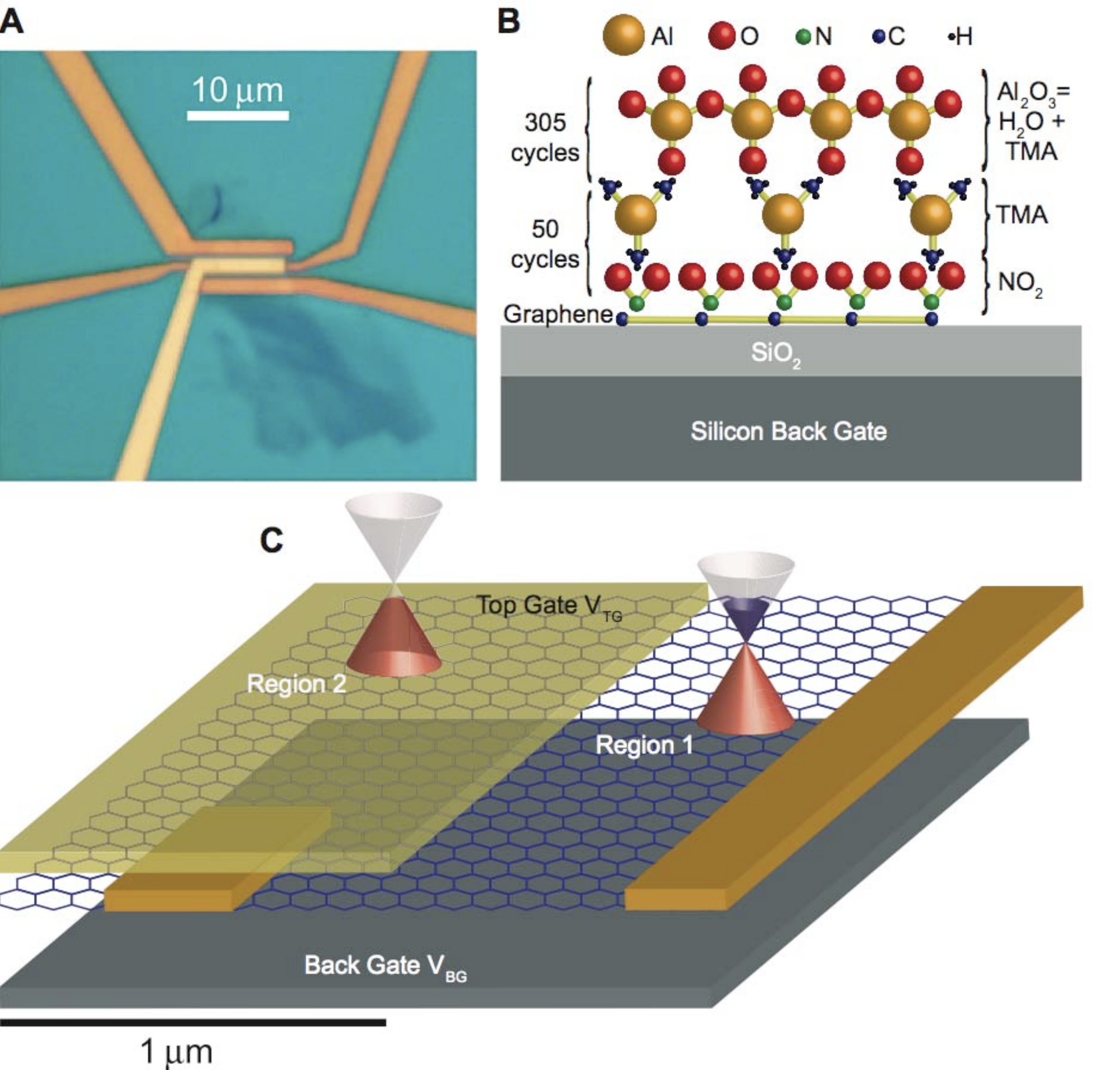}
\caption{\footnotesize{
(\textbf{A}) Optical micrograph of a device similar to
the one measured. Metallic contacts and top gate appear in orange
and yellow, respectively. Darker regions below the contacts are
thicker graphite from which the contacted single layer of graphene
extends. (\textbf{B}) Illustration of the oxide deposition process.
A non-covalent functionalization layer is first deposited using
NO$_2$ and TMA (50 cycles) and $\alox$ is then grown by atomic layer
deposition using H$_2$O-TMA (305 cycles yielding $\sim$30~nm
thickness). (\textbf{C}) Schematic diagram of the device measured in
this experiment.
}}
\end{figure}

Graphene sheets are prepared via mechanical exfoliation using a
method similar to that used in Ref.~10. Graphite flakes are
deposited on 300~nm of SiO$_2$ on a degenerately doped Si substrate.
Inspection with an optical microscope allows potential single-layer
regions of graphene to be identified by a characteristic coloration
that arises from thin-film interference. These micron-scale regions
are contacted with thermally evaporated Ti/Au (5/40~nm), and
patterned using electron-beam lithography. Next, a $\sim$ 30~nm
layer of oxide is deposited atop the entire substrate. As
illustrated~(Fig.~1B), the oxide consists of two parts: a
non-convalent functionalization layer (NCFL)  and $\alox$. This
deposition technique is based on a recipe successfully applied to
carbon nanotubes~\cite{Farmer06}. The NCFL serves two purposes. One
is to create a non-interacting layer between the graphene and the
$\alox$ and the other is to obtain a layer that is catalytically
suitable for the formation of $\alox$ by atomic layer deposition
(ALD). The NCFL is synthesized by 50 pulsed cycles of NO$_2$ and
trimethylaluminum (TMA) at room temperature inside an ALD reactor.
Next, 5 cycles of H$_2$O-TMA are applied at room temperature to
prevent desorption of the NCFL. Finally, $\alox$ is grown at
$225^\circ$C with 300 H$_2$O-TMA ALD cycles. To complete the device,
a second step of electron-beam lithography defines a local top gate
(5/40~nm Ti/Au) covering a region of the device that includes one of
the metallic contacts.

A completed device, similar in design to that shown in the optical
image in Fig.~1A, was cooled in a $\He3$ refrigerator and
characterized at temperatures $T$ of $250~\mathrm{mK}$ and
$4.2~\mathrm{K}$. Differential resistance $R=dV/dI$, where $I$ is
the current and $V$ the source-drain voltage, was measured by
standard lock-in techniques with a current bias of
$1~(10)~\mathrm{nA}_{\mathrm{rms}}$ at $95~\mathrm{Hz}$ for
$T=250~\mathrm{mK}~(4.2~\mathrm{K})$. The voltage across two
contacts on the device, one outside the top-gate region and one
underneath the top gate,  was measured in a four-wire configuration,
eliminating series resistance of the cryostat lines. A schematic of
the device is shown in Fig.~1C.

The differential resistance $R$ as a function of back-gate voltage
$\VBG$ and top-gate voltage $\VTG$ at $B=0$ (Fig.~2A), demonstrates
independent control of carrier type and density in the two regions.
This two-dimensional (2D) plot reveals a skewed, cross-like pattern
that separates the space of top-gate and back-gate voltages into
four quadrants of well-defined carrier type in the two regions of
the sample. The horizontal (diagonal) ridge corresponds to charge
neutrality,~i.e.,~the Dirac point, in region 1 (2). The slope of the
charge-neutral line in region 2, along with the known distances to
the top gate and back gate, gives a dielectric constant $\kappa \sim
6$ for the functionalized $\alox$. The center of the cross at
$(\VTG,\VBG)\sim (-0.2~\mathrm{V},-2.5~\mathrm{V})$ corresponds to
charge neutrality across the entire graphene sample. Its proximity
to the origin of gate voltages demonstrates that the functionalized
oxide does not chemically dope the graphene significantly.

\begin{figure}
\label{fig2}
\includegraphics[width=3in]{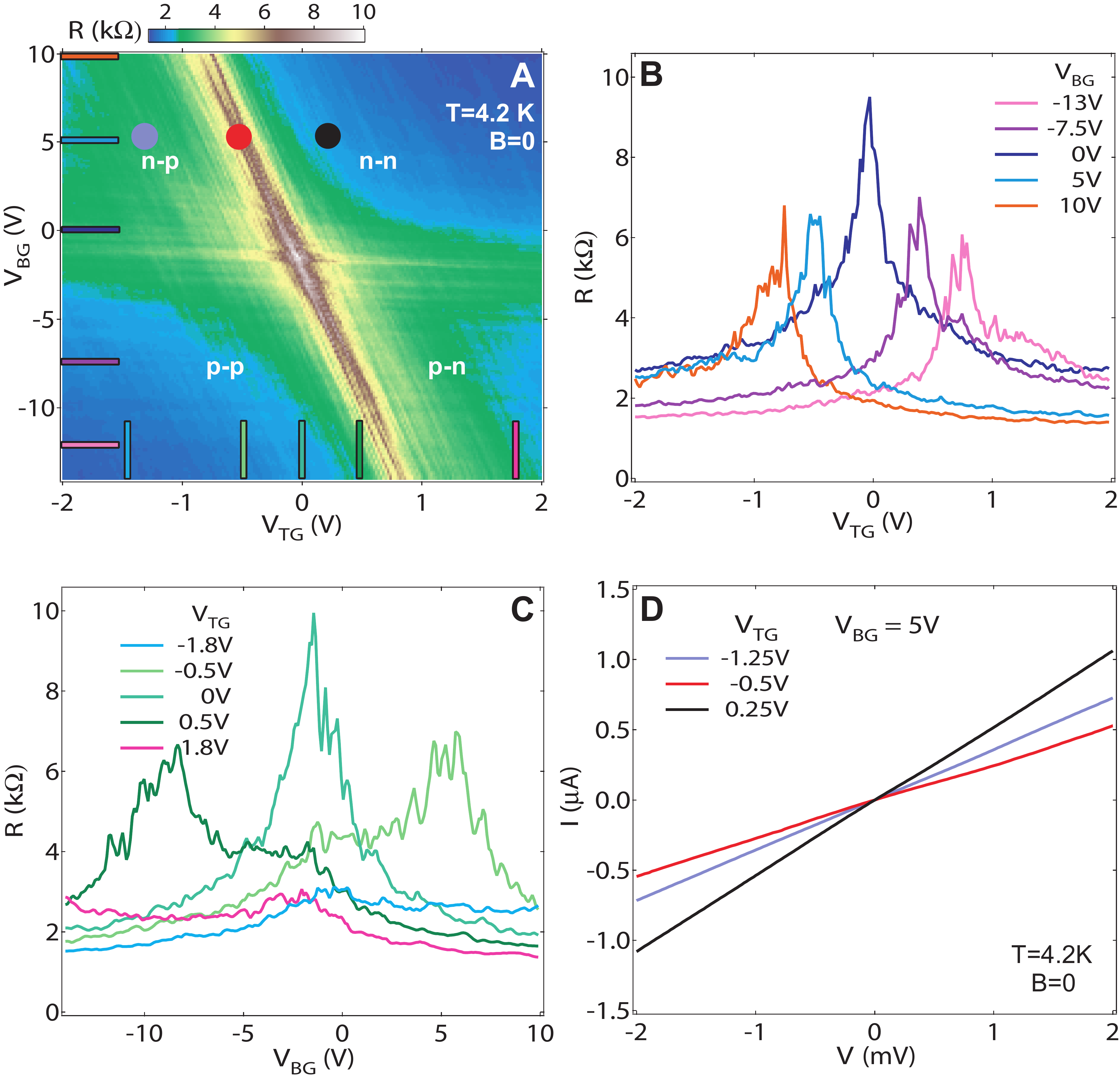}
\caption{\footnotesize{
(\textbf{A}) Two-terminal differential resistance $R$
as a function of the top-gate voltage $\VTG$ and back-gate voltage
$\VBG$ at $B=0$ and $T=4.2~\mathrm{K}$, demonstrating independent
control of carrier type and density in regions 1 and 2. Labels in
each of the four quadrants indicate the carrier type (first letter
indicates carrier type in region 1). (\textbf{B} and \textbf{C})
Horizontal (Vertical) slices at $\VBG~(\VTG)$ settings corresponding
to the colored lines superimposed on Fig.~2A. (\textbf{D}) $I$-$V$
curves at the gate voltage settings corresponding to the solid
circles in Fig.~2A are representative of the linear characteristics
observed everywhere in the plane of gate voltages.
}}
\end{figure}

Slices through the 2D conductance plot at fixed $\VTG$ are shown in
Fig.~2C. The slice at $\VTG=0$ shows a single peak commonly observed
in devices with only a global back
gate~\cite{Novoselov04,Novoselov05,Zhang05,Heersche07}. Using a
Drude model away from the charge-neutrality region, mobility is
estimated at $\sim
7000~\mathrm{cm}^2/\mathrm{Vs}$~\cite{Novoselov04}. The peak width,
height, and back-gate position are consistent with single-layer
graphene~\cite{Novoselov05,Zhang05,Heersche07} and provides evidence
that the electronic structure and degree of disorder of the graphene
is not strongly affected by the oxide. Slices at finite $|\VTG|$
reveal a doubly-peaked structure. The weaker peak, which remains
near $\VBG \sim -2.5~\mathrm{V}$ at all $\VTG$, corresponds to the
Dirac point of region 1.  The stronger peak, which moves linearly
with $\VTG$, is the Dirac point for region 2. The difference in peak
heights is a consequence of the different aspect ratios of regions 1
and 2. Horizontal slices at fixed $\VBG$ corresponding to the
horizontal lines in Fig.~2A are shown in Fig.~2B. These slices show
a single peak corresponding to the Dirac point of region 2. This
peak becomes asymmetric away from the charge-neutrality point in
region 1. We note that the $\VBG$ dependence of the asymmetry is
opposite to that observed in Ref.~9, where the asymmetry is studied
in greater detail. The changing background resistance results from
the different density in region 1 at each $\VBG$ setting.
Current-voltage ($I$-$V$) characteristics, measured throughout the
$(\VTG,\VBG)$ plane, show no sign of rectification in any of the
four quadrants or at either of the charge-neutral boundaries between
quadrants~(Fig.~2D), as expected for reflectionless (``Klein'')
tunneling at the $p$-$n$ interface~\cite{Katsnelson06,Cheianov06}.

In the QH regime at large $B$, the Dirac-like energy spectrum of
graphene gives rise to a characteristic series of QH plateaus in
conductance, reflecting the presence of a zero-energy Landau level,
that includes only odd multiples of $2~\gq$ (that is, 2, 6,
10,...$~\times \gq$) for uniform carrier density in the
sheet~\cite{Gusynin05,Abanin06,Peres06}. These plateaus can be
understood in terms of an odd number of QH edge states (including a
zero-energy edge state) at the edge of the sheet, circulating in a
direction determined by the direction of $B$ and the carrier type.
The situation is somewhat more complicated when varying local
density and carrier type across the sample.

A 2D color plot of differential conductance $g=1/R$ as a function of
$\VBG$ and $\VTG$ at $B=4~\mathrm{T}$ is shown in Fig.~3A. A
vertical slice at $\VTG=0$ through the $p$-$p$ and $n$-$n$
quadrants~(Fig.~3B) reveals conductance plateaus at 2, 6, and 10
$\gq$ in both quadrants, demonstrating that the sample is
single-layer and that the oxide does not significantly distort the
Dirac spectrum.

QH features are investigated for differing filling factors $\nuone$
and $\nutwo$ in regions 1 and 2 of the graphene sheet. A horizontal
slice through Fig.~3A at filling factor $\nuone=6$ is shown in
Fig.~3C. Starting from the $n$-$n$ quadrant, plateaus are observed
at $6~\gq$ and $2~\gq$ at top-gate voltages corresponding to filling
factors $\nutwo=6$ and $2$, respectively. Crossing over to the
$n$-$p$ quadrant by further decreasing $\VTG$, a new plateau at
$3/2~\gq$ appears for $\nutwo=-2$. In the $\nutwo=-6$ region, no
clear QH plateau is observed. Another horizontal slice at
$\nuone=~2$ shows 2~$\gq$ plateaus at both $\nutwo=6$ and $2$ (see
Fig.~3D). Crossing into the $n$-$p$ quadrant, the conductance
exhibits QH plateaus at $1~\gq$ for $\nutwo=-2$ and near~$3/2~\gq$
for $\nutwo=-6$.

For $\nuone$ and $\nutwo$ of the same sign ($n$-$n$ or $p$-$p$), the
observed conductance plateaus follow
\begin{equation}
g = \mathrm{min}(|\nuone|, |\nutwo|)\times \gq.
\end{equation}
This relation suggests that the edge states common to both regions
propagate from source to drain while the remaining $|\nuone-\nutwo|$
edge states in the region of highest absolute filling factor
circulate internally within that region and do not contribute to the
conductance. This picture is consistent with known results on
conventional 2D electron gas systems with inhomogeneous electron
density~\cite{Syphers85,Haug88,Washburn88}.

\begin{figure}
\label{fig3}
\includegraphics[width=3in]{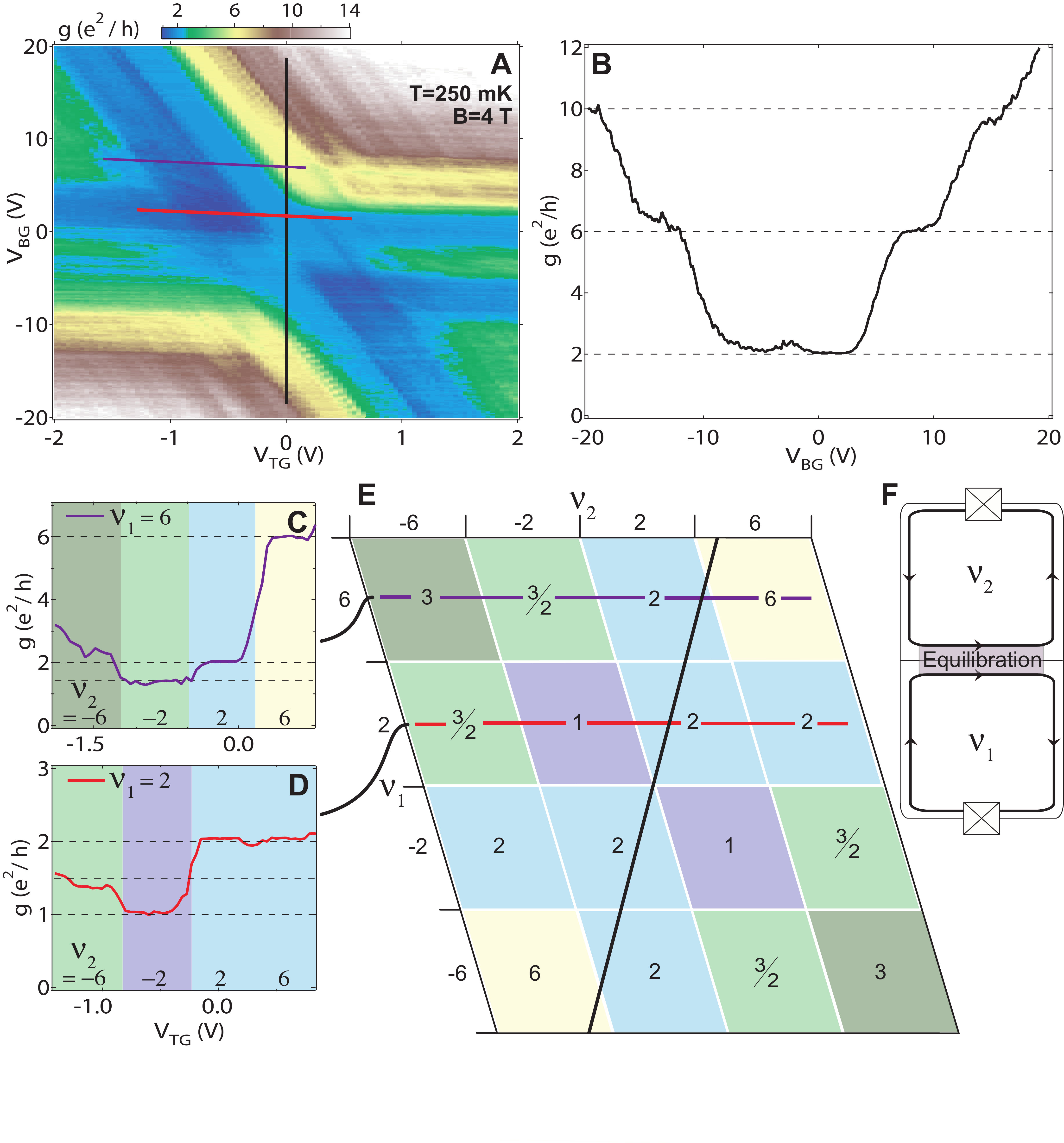}
\caption{\footnotesize{
(\textbf{A}) Differential conductance $g$ as a function
of $\VTG$ and $\VBG$ at $B=4~\mathrm{T}$ and $T=250~\mathrm{mK}$.
(\textbf{B}) Vertical slice at $\VTG=0$, traversing $p$-$p$ and
$n$-$n$ quadrants. Plateaus are observed at $2~\gq$ and $6~\gq$, the
quantum Hall signature of single-layer graphene. (\textbf{C})
Horizontal slice at $\nuone=6$ showing conductance plateaus at 6,~2
and $3/2~\gq$. (\textbf{D}) Horizontal slice at $\nutwo$ showing QH
plateaus at $2$, $1$ and $3/2~\gq$. (\textbf{E}) Table of
conductance plateau values as a function of filling factors
calculated using Eqs.~1 and 2. Black, purple and red lines
correspond to slices in (\textbf{B}), (\textbf{C}) and (\textbf{D}),
respectively. (\textbf{F}) Schematic of counter-circulating edge
states at filling factors
$\nuone=-\nutwo=2$.
}}
\end{figure}

Recent theory~\cite{Abanin07} addresses QH transport for filling
factors with opposite sign in regions 1 and 2 ($n$-$p$ and $p$-$n$).
In this case, counter-circulating edge states in the two regions
travel in the same direction along the $p$-$n$ interface (Fig.~3F),
which presumably facilitates mode mixing between parallel-traveling
edge states. For the case of complete mode-mixing---that is, when
current entering the junction region becomes uniformly distributed
among the $|\nuone|+|\nutwo|$ parallel-traveling modes---quantized
plateaus are expected~\cite{Abanin07} at values
\begin{equation}
g=\frac{|\nuone||\nutwo|}{|\nuone|+|\nutwo|}\times \gq.
\end{equation}
A table of the conductance plateau values given by Eqs.~1 and 2 is
shown in Fig.~3E. Plateau values at $1~\gq$ for $\nuone=-\nutwo=2$
and at $3/2~\gq$ for $\nuone=6$ and $\nutwo =-2$ are observed in
experiment. Notably, the $3/2~\gq$ plateau suggests uniform mixing
among four edge stages (three from region 1 and one from region 2).
All observed conductance plateaus are also seen at $T=4~\mathrm{K}$
and for $B$ in the range 4 to 8~T.

We do find some departures between the experimental data and Eqs.~1
and 2, as represented in the grid of Fig.~3E. For instance, the
plateau near 3/2 $\gq$ in Fig.~3D is seen at a value of $\sim$
$1.4~\gq$ and no clear plateau at $3~\gq$ is observed for
$\nuone=-\nutwo = 6$. We speculate that the conductance in these
regions being lower than their expected values is an indication of
incomplete mode mixing. We also observe an unexpected peak in
conductance at a region in gate voltage between the two $1~\gq$
plateaus at $\nuone=\pm\nutwo=2$. This rise in conductance is
clearly seen for $|\VTG|$ values between $\sim$ 1 and 2~V and $\VBG$
values between $\sim$ -5 and -2~V. This may result from the possible
existence of puddles of electrons and holes near the
charge-neutrality points of regions 1 and 2, as previously
suggested~\cite{Hwang06b}.

We thank L.~S. Levitov, D.~A. Abanin, C.~H. Lewenkopf, and P.
Jarillo-Herrero for useful discussions. We thank Z. Chen at IBM T.
J. Watson Research Center for suggesting the NO$_2$
functionalization process and D. Monsma for assistance in
implementing it. Research supported in part by INDEX, an NRI Center,
and by the Harvard NSEC.

\end{document}